\def\bbbz{{\bf Z}}
\def\A{{\cal A}}
\def\F{{\cal F}}
\def\L{{\cal L}}

\documentclass[proceedings,12pt]{JHEP}
\conference{Nonperturbative Quantum Effects 2000}

\usepackage{epsfig}

\title{\Large\bf The Low Energy Dynamics of Non-BPS Branes}

\author{N.D. Lambert${}^1$ and I. Sachs${}^2$
\\ ${}^1$Dept. of Mathematics\\ King's College London\\
The Strand,  London\\  WC2R 2LS,  UK.\\
${}^2$Theoretische Physik\\
Ludwig-Maximilians Universit\"at\\
Theresienstrasse 27 \\
80333 Munich\\ Germany.\thanks{KCL-TH-00-58}}

\abstract{In this talk we will discuss the low energy dynamics of 
non-BPS branes constructed as stable brane/anti-brane pairs at an
orbifold. In particular we will determine the effective field
theory  and compare its predictions with those of the full open string
theory. While the position and vector degrees of freedom of the
branes have the familar form found in supersymmetric gauge theories, 
the massless modes originating in the tachyonic sector display novel 
non-commuting flat directions. We will show that
these flat directions persist to all orders in $\alpha'$. Finally
we will briefly discuss the open string loop corrections.}

\begin{document}

\section{Introduction}

In the past two years Sen has pioneered the study of stable non-BPS branes
in string theory. Here we consider stable non-BPS states 
constructed as brane/anti-brane pairs on an orbifold that 
removes any tachyonic instabilities. The existence of such states 
first appeared as a consequence of various string dualities.
A natural question that arises with stable non-BPS branes is to obtain
their low energy effective Lagrangian.
Indeed there are several motivations for this study.  

The first is that in the
case of BPS branes,  most of their dynamics can be
readily understood from the more familiar dynamics of the quantum gauge
field theory living on their worldvolume. In fact only the lowest order
term in an $\alpha'$ expansion of the effective Lagrangian is usually needed
and this is none other than that of a maximally supersymmetric 
Yang-Mills theory.
Therefore  one might hope that by deriving the low energy effective theory 
for  non-BPS branes 
one would find a simpler and more familiar description of their dynamics.
On the other hand the lack of supersymmetry leads one to question how
appropriate the low energy field theory approximation is.

Branes, most recently through the AdS/CFT
correspondence, have allowed for radical and new approaches to the study
of the strong coupling limit of supersymmetric field theories. One might
think that non-BPS branes will open an analogous study
for non-supersymmetric gauge theories. However there is a large obstacle
to this approach. In the AdS/CFT correspondence a crucial role is played
by the low energy supergravity solution for a large number of branes. 
As is well known, for BPS-branes there is a 
no-force condition and hence there
is no limit to the number and location of the branes. However it was shown in
\cite{GS} that there is a one loop 
repulsive force between two parallel stable non-BPS
branes, except at the point in their moduli space where they become
unstable. Hence an important question that
remains is whether or not large numbers of non-BPS branes can somehow be 
prevented from flying apart so that a supergravity solution can be
found (see \cite{EP,L,Oz,DiVecchia} for 
an interesting discussion of this point).

It is also very natural to suppose that non-BPS branes have an important
role to play in phenomenological string models, since they have no worldvolume
supersymmetry (although supersymmetry is preserved in the bulk). 
To our knowledge
however, there has not yet  
appeared any attempts in the literature along these lines.

In this talk we will review our study of the effective field theory for
stable non-BPS branes \cite{LS,LS'} and   
also briefly report on some work in progress \cite{LSW}.
No attempt is made here to survey the literature as a whole  
or to provide explicit technical details, which are available in the
original papers. Some more details may also be found in the talk of
Janssen and Meessen in these proceedings \cite{JM}. 
Our plan is as follows.
First we will provide
a brief review of non-BPS branes. 
Secondly we will obtain the classical low energy effective
Lagrangian for a stack of parallel non-BPS branes, to all orders 
in $\alpha'$.
We will do this for the specific case of a non-BPS 3-brane at the critical
orbifold radius, beyond which it becomes unstable. 
For a general radius of the orbifold 
the full effective Lagrangian is unknown and obtaining it appears to be a
very challenging problem, although the lowest order term in $\alpha'$ is
known \cite{LS}.
Thirdly we will provide a description of the classical moduli space of
vacua and some of the topologically stable soliton states. Finally we will
discuss the one loop quantum corrections to the effective
Lagrangian. This will be done in  in two ways; from the lowest order 
(renormalisable) term in
the effective quantum field theory and also from the full open string theory.
This allows us to compare the two predictions and test to what extent the
quantum field theory approximation captures the full dynamics of the string 
theory.

\section{Review of Stable Non-BPS Branes Wrapped on $T^{4}/{\bbbz}_2$.}

We start by briefly recalling the key features
non-BPS D$p$-branes as presented in the reviews articles 
\cite{Sen9904207,Gaberdiel}. 
A non-BPS D$p$-brane (which we denote as a ${\tilde D}p$-brane) 
of type IIA (type IIB) string
theory   can be defined by starting with a D$p$-brane/anti-D$p$-brane
pair of type IIB (type IIA) string theory. One then mods-out the
theory by $(-1)^{F_L}$, which changes the sign of the left-moving
spacetime Fermions. In the bulk this takes type IIB to
type IIA string theory (or vice-versa). In terms of the open strings
ending on the D$p$-branes this removes half of the Chan-Paton indices.
In this way we see that  a ${\tilde D}p$-brane has  
two types of open strings ending on it, labelled by the Chan-Paton
indices $I$ and $\sigma_1$. Contrary to BPS D$p$-branes one has $p$ even
in type IIB string theory and $p$ odd in type IIA string theory.
The $I$-sector strings are precisely the
same as for BPS D-branes and therefore their low energy modes form
a maximally supersymmetric vector multiplet in $(p+1)$ dimensions.
The $\sigma_1$-sector strings come with the opposite projection 
under $(-1)^F$.
After the GSO projection, the level one sector modes of
the NS-sector are projected out while the tachyonic ground state survives.
Therefore the lightest modes in the $\sigma_1$-sector 
consist of real tachyonic scalar with mass-squared
$-1/2\alpha'$ and a sixteen-component massless Fermion. 

Clearly this system is unstable due to the tachyon. However by wrapping
the brane around an orbifold, for example we will take 
$T^{4}/\bbbz_{2}$ with coordinates  $i = 6,\ldots,9$ and radii $R_i$, 
the tachyonic modes which are even under the orbifold action 
$g:x^i\leftrightarrow -x^i$ are projected out. Thus only the 
tachonic modes with odd momentum around the orbifold survive. Since
a momentum mode has mass-squared
\begin{equation}
m^2 = -{1\over 2\alpha'}+ \sum_{i=6}^9 \left(n_i\over R_i\right)^2\ ,
\end{equation}
we see that below a critical radius $R_c = \sqrt{2\alpha'}$ the 
surviving states are non-tachyonic. 
In particular the four lightest tachyon scalars surviving 
the orbifold are given by 
\begin{equation}
        T^{i}=\frac{\chi^{i}(x)}{2i}\left(e^{i\vec{\omega}^{i}\cdot
        \vec{X}}-e^{-i\vec{\omega}^{i}\cdot\vec{X}}\right),
\label{tach}
\end{equation}
where $\vec{\omega}^{i} =  {(R_i)^{-1}\vec{e}^{i}}$. In this way we
obtain a stable ${\tilde D}p$-brane.

On the other
hand, the effect of the orbifold on the Bosons of the $I$-sector strings
are the same as for a D$p$-brane at an orbifold. Therefore, in the
case of a ${\tilde D}7$-brane wrapped on $T^{4}/\bbbz_{2}$, 
which we will be most interested in here\footnote{Note that this is
T-dual to the case considered in \cite{LS}.}, 
the $N=4$ vector multiplet is reduced to an $N=2$ multiplet. 
This contains two
scalars $X^I$, $I=4,5$, that represent the fluctuations of the brane
in the $x^4$ and $x^5$ directions, and a vector $A_\mu$, $\mu=0,1,2,3$. 
For the Fermions the orbifold selects out  one chirality under
$\Gamma^{6789}$ for the $I$-sector and the 
opposite for the $\sigma_1$-sector.
This is the same Fermionic content as four-dimensional $N=4$ 
super-Yang-Mills.

For most of this talk we will restrict our attention to oribifolds at 
the critical radius $R_i=\sqrt{2\alpha'}$ where the lightest tachyon modes 
are massless.  
We will only use the Einstein summation convention
for the full ten-dimensional indices $m,n=0,1,2...,9$ and the
four-dimensional indices $\mu,\nu=0,1,2,3$. 
All other sums (e.g. those over the
$i,j$ and $I,J$ indices) will be explicitly written.

\section{The Effective Lagrangian for Parallel non-BPS Branes}

In general it is a very hard problem to analyse the effect of
relevant vertex operators on the string worldsheet.
However, at the the critical radius the tachyon vertex operators
are marginal. Indeed at this radius it is possible to
Fermionise,
and then re-Bosonise, these vertex operators into the form of 
Wilson lines \cite{Sen9808141,Sen0003124} as follows. One starts by expressing
the open string fields $X^i$  and $\psi^i$ in terms of the boundary values of 
left and right-handed closed string fields: $X^i = 2X_L^i = 2X^i_R$
and $\psi^i = \psi^i_L=\psi_R^i$. Next one Fermionises these closed
string fields as
\begin{equation}
e^{i\sqrt{{2}/{\alpha'}}X_{L/R}^i} = \frac{1}{2}(\xi_{L/R}^i +
i\eta_{L/R}^i)\otimes \Gamma^i\ .
\end{equation}
Here the $\Gamma^i$ matrices form a representation
of the $Spin(4)$ Clifford algebra. Furthermore the other worldsheet
Fermions are now tensored with a $\Gamma^5$, e.g. 
$\psi^i_{L/R}\rightarrow \psi^i_{L/R}\otimes \Gamma^5 $. 
These $\Gamma$-matrices  arise as co-cycles and
ensure that the worldsheet fields obey the correct statistics,
i.e. that $\xi^i_{L/R}$ and $\eta^i_{L/R}$ commute with worldsheet
Bosons but anti-commute with worldsheet Fermions. Finally one can
re-Bosonise by taking
\begin{equation}
\frac{1}{2}(\xi_{L/R}^i + 
i\psi_{L/R}^i) = e^{i\sqrt{{2}/{\alpha'}}{\tilde X}_{L/R}^i} \otimes {\tilde \Gamma}^i\ ,
\end{equation}
where the $\tilde \Gamma$-matrices are another representation of the
$Spin(4)$ Clifford algebra. In this way we find an equivalent
description
of the worldsheet fields in terms of $X^m,\psi^m,{\tilde X}^i,\eta^i$,
$m=0,1,2,...,5$, $i=6,7,8,9$. After this change of variables 
one can see that the tachyon vertex
operators corresponding to the states \ref{tach} 
are mapped into  Wilson lines but with the enlarged
Chan-Paton factor $\chi^i\otimes\sigma_1\otimes i\Gamma^5\Gamma^i$.
Here the $\sigma_1$ represents the fact that the tachyon comes from
the $\sigma_1$-sector and the $i\Gamma^5\Gamma^i$ factor comes from the
co-cycles.

The analysis for all the degrees of freedom is most 
easily  presented after T-dualising in the $x^4$ and 
$x^5$-directions. In this case
the positions of the branes are also given by Wilson lines. 
Thus, with the aid of these new variables, we now 
consider $N$ ${\tilde D}9$-branes wrapped around $T^{4}/\bbbz_{2}\times 
T^{2}$, which are described by the Wilson line $\A_{m}$, $m=0,1,2,...,9$
\cite{LS'}
\begin{equation}
        \A_{m}=\cases{A_{\mu}\otimes I\otimes I;\quad m=0,\cdots,3\cr
        \phi^{I}\otimes I\otimes I;\quad m=4,5\cr
        \chi^{i}\otimes \sigma^{1}\otimes i\Gamma^5\Gamma^{i}
        ;\quad m=6,\cdots,9}
        \label{BI1}
\end{equation}
where the fields $A_\mu,\phi^I,\chi^i$ all take values in the Lie
algebra of $U(N)$. 

We can now compute the effective Lagrangian for a ${\tilde D}$9-brane
within the existing formalism for Wilson lines
(for example see \cite{Tseytlin} and the references therein). 
In general the effective Lagrangian will receive 
corrections to all orders in $\alpha'$ 
and $\F_{mn}=\partial_m \A_n - \partial_n \A_m
-ig[\A_m,\A_n]$ and we are unable to find an analytic 
expression. A simplification occurs if we neglect terms of the 
form $D_{(m}\F_{n)p}$, in which case we find \cite{Tseytlin}
\begin{equation}
  \L_B=c_0\hbox{STr}\sqrt{\det(\delta_{mn}+2\pi\alpha'\F_{mn})}\ .
  \label{BI5}
\end{equation}
where $c_0$ is a constant and $\hbox{STr}$ is the symmeterised trace. 
Note that the  determinant here is taken over $m,n$ indices and the
trace is over the Chan-Paton and co-cycle factors in \ref{BI1}. 

Next we may T-dualise and take the limit where $x^4$ and $x^5$ dimensions 
decompactify to obtain the effective Lagrangian for 
various ${\tilde D}p$-branes, viewed as ${\tilde D}(p+4)$-branes
wrapped  over the orbifold. 
Here we will concentrate on a
${\tilde D}3$-branes.
The lowest non-trivial term of \ref{BI5} gives a renormalisable
effective Lagrangian
\begin{eqnarray}
 \label{BI7}
\L_B &=& c_0' {\rm Tr} \left(
\frac{1}{4}F_{\mu\nu}F^{\mu\nu} +\frac{1}{2}\sum_I D_\mu \phi^I D^\mu \phi^I
\right.\nonumber\\
&&\quad\left.+\frac{1}{2}\sum_i D_\mu \chi^i D^\mu \chi^i - V \right)\ ,\nonumber\\
V&=&\frac{g^2}{4}\sum_{I,J}([\phi^I,\phi^J])^2
+\frac{g^2}{2}\sum_{I,j}([\phi^I,\chi^j])^2 \nonumber\\
&&-\frac{g^2}{4}\sum_{i\ne j}(\{\chi^i,\chi^j\})^2\ ,
\end{eqnarray}
where $c_0'$ is a constant and $D_\mu = \partial_\mu - ig[A_\mu,\ ]$.
Here we have dropped a constant 
term in $V$ and the trace is now only over
the group indices $a,b$. 

The lowest order terms in the low 
energy effective Lagrangian can also be obtained by taking
the $\alpha'\rightarrow 0$ limit of the disk diagrams 
for the open string S-matrix \cite{LS}.
The Bosonic part of the resulting effective action agrees precisely
with \ref{BI7}. 
However in addition the couplings of the  (massless) Fermionic fields on a
${\tilde D}p$-brane can be determined \cite{LS}. 
If we denote the Fermions arising the $I$-sector and $\sigma_1$-sectors 
by $\lambda$ and $\psi$ respectively then, to lowest order, 
the total effective Lagrangian
has the form $\L = \L_B + \L_F$ where
\begin{eqnarray}
\L_F &=& c_0'{\rm Tr} \left(i\bar\lambda \gamma^\mu D_\mu\lambda
+i\bar\psi\gamma^\mu D_\mu\psi \right. \nonumber\\
&&\left.+ g\bar\lambda\gamma^I[\phi^I,\lambda]
+ g\bar\psi\gamma^I[\phi^I,\psi]\right)\ .
\end{eqnarray}
Note that there are no Yukawa interactions involving the four scalars
$\chi^i$. 
Thus, even though at the critical radius where we have exactly the same field
content as a maximal Yang-Mills multiplet, the effective Lagrangian is not
supersymmetric. Therefore one can expect that there will be
non-trivial loop corrections in both the string theory and field 
theory approximation.

The string S-matrix also allows one to work away from the critical radius,
where the modes $\chi^i$ have a mass $m^2_i = 1/R_i^2 - 1/2\alpha'$. In this
case the only change to the Lagrangian  is the appearance of a 
mass term
$\frac{1}{2}m_i^2\chi_i^2$ in $V$ \cite{LS}. 
Note that if the ${\tilde D}p$-brane
is unstable, so that at least one $m_i^2$ is negative, then the
potential $V$ is unbounded from below since we may make the 
corresponding $\chi^i$ as large as we wish with the other fields vanishing
and $V$ will then be arbitrarily negative. 
Therefore the lowest order
(renormalisable) term in the effective Lagrangian is not a valid
description for unstable ${\tilde D}p$-brane. 

\section{The Classical Vacua and BPS States of Non-BPS Branes}

Perhaps the most notable feature of the effective Lagrangian \ref{BI7}
is that the classical vacuum moduli space, given by the space of
constant scalars, has non-Abelian flat directions. Indeed we can see
that the condition for constant $\chi^i$ and $\phi^I$ scalar 
vertex operators to be
marginal is the zero curvature constaint $[\A_m,\A_n]=0$ or, more explicitly,
\begin{equation}
[\phi^I,\phi^J] = [\phi^I,\chi^i] = \{\chi^i,\chi^j\}=0\ ,
\end{equation}
for $i\ne j$ and to all orders in $\alpha'$. 
Of course since there is
no supersymmetry we do not expect the situation to persist once string
loops are taken into account.

We also note  that, in a vacuum parameterised by anti-commuting vev's for
$\chi^i$, the gauge group can be broken down to a single $U(1)$
representing the centre of mass, rather
than $U(1)^N$ as is the case for vacua in which the scalars $\phi^I$
have a non-vanishing vev. In 
addition the components of $\phi^I$ which are not in this $U(1)$ will
become massive. Thus, since the $\phi^I$ describe the separation of
the branes,  we see that in these vacua the non-BPS branes
are classically bound together. Clearly if $m_i\ne0$
then the $\chi^i$ vev's must vanish.

Let us now mention some of the soliton solutions on the 
${\tilde D}$3-brane worldvolume. 
Although one can find several types of BPS states of the Lagrangian
\ref{BI7} \cite{LS'}, here we will concentrate  on the monopoles. If
we turn on just one scalar, either $\chi^i$ or $\phi^I$, then there
is no potential to all orders in $\alpha'$ and the Bosonic Lagrangian 
is identical to that obtained from a maximally 
supersymmetric D3-brane (in the case of only one non-zero scalar). 
It then follows that
the BPS monpoles given by the first order equations $B_a = D_a\phi^I$
or $B_a =D_a \chi^i$  ($a=1,2,3$), where 
$B_a = \frac{1}{2}\epsilon_{abc}F^{bc}$ is the magnetic field,   
are absolute minimum energy configurations for
a fixed topological index given by the magnetic charge. Therefore, even
without supersymmetry, we expect these states to be stable.

It is well known that on a D3-brane a 
monopole configuration involving the scalar $\phi^I$
corresponds to a D1-brane suspended between two D3-branes along
the $x^I$ direction.  A similar interpretation occurs here expect that,
since we are in type IIA string theory, it must be a 
${\tilde D}$1-brane
suspended between two ${\tilde D}$3-branes.
On the other hand a tachyon
vev $\chi^i=1$ corresponds to deforming a single  wrapped 
${\tilde D}7$-brane into a
D6-brane and an anti-D6-brane wrapped over the $x^6,x^7,x^8$
directions but sitting at opposite ends of the
orifold \cite{Sen9904207}. 
Therefore a monopole soliton
involving $\chi^i$ 
can be pictured as two wrapped ${\tilde D}$7-branes 
at the monopole core, but at infinity the two ${\tilde D}$7-branes 
split up into 
D6-brane/anti-D6-brane and anti-D6-brane/D6-brane pairs at opposite
ends of the orbifold.
These states again correspond to ${\tilde D}$1-branes but this time
stretched along an orbifold direction $x^i$. 

Now consider what happens if we move away from the critical radius. 
In this case the
a mass term $\frac{1}{2}m_i^2\chi_i^2$ appears in the Lagrangian and so 
the $\chi^i$ monopoles no longer exist. This can be understood because a
${\tilde D}$3-brane is stable and hence 
the $m_i^2$ are positive only if the orbifold radii are
larger than the critical radius (note that we are in the T-dual
picture compared to section two). 
However a ${\tilde D}$1-brane stretched along
an orbifold direction is stable only if the radius is less than critical.

\section{Quantum Corrections}

Finally we would like to discuss the quantum corrections to the classical
effective Lagrangian described above. We will do this in two ways. First we
will calculate the one loop
quantum corrections in the full string theory \cite{LSW}. 
Second we 
will consider the renormalisable quantum field theory given by the lowest
order terms in the effective Lagrangian to test how well it reproduces
the string theory results. For
simplicity
we consider the case of two ${\tilde D}$3-branes.

\subsection{Open String Theory}

It was shown in \cite{GS} that away from the critical radius, where
they are stable, there is a repulsive force between
two parallel ${\tilde D}p$-branes at any separation. 
At the critical radius the force vanishes because there is a spurious 
Bose-Fermi degeneracy, at
all levels in the open string theory, and so the one loop effective
potential (which is the negative of the
partition function) vanishes. 

\EPSFIGURE{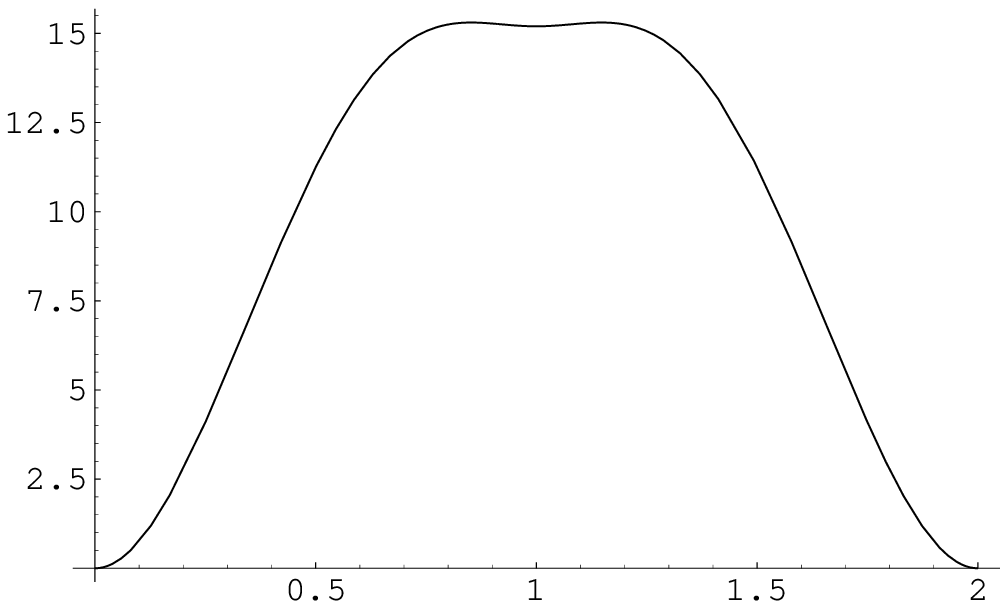,width=3cm}{$V_{eff}$ as a function of $v_0$}
We have seen above that at the critical radius there a new branches
of the classical vacuum moduli space associated with non-Abelian
flat directions. Here we will report on some work in progress to
determine the partition function in these branches \cite{LSW}. 
There are two cases that are
of interest. Either $<\chi^9> = v_0I$ or $<\chi^9> = v_3\sigma_3$, 
where $\sigma_3 \in  su(2)$. 
More complicated vacua, with more than
one tachyon non-vanishing can also be considered but the calculation becomes
increasingly complicated and we do not expect the general form to
change. The plots for the effective potential 
for these two cases are given in figures 1 and 2
respectively. 
\EPSFIGURE{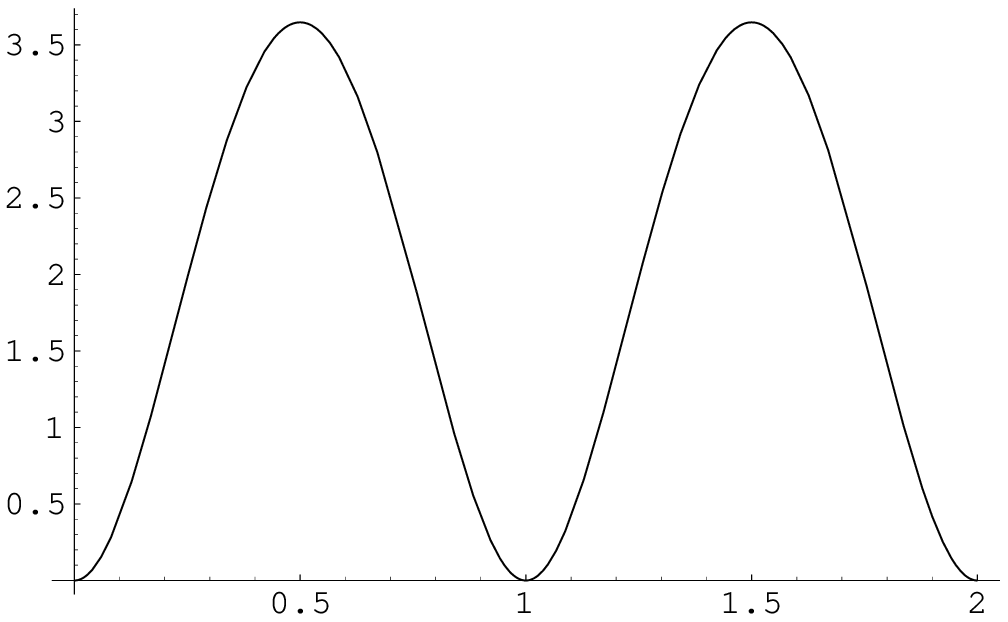,width=3cm}{$V_{eff}$ as a function of $v_3$}
Note that these plots are periodic in $v_0$ with period 2 and 
periodic in $v_3$ with period 1. In both of these cases there
is a divergence in the integral over the open string modular
parameter. These are related to the existence of 
tadpoles diagrams for the closed string fields via the
Fischler-Susskind mechanism \cite{LSW}. 

Turning on a vev $<\chi^9>=1$ on wrapped ${\tilde D}7$-branes
corresponds to deforming all of them  into  D6-brane/anti-D6-brane
pairs at opposite ends of the  the orbifold, i.e. the D6-branes are
at one end point and the anti-D6-branes at the other. 
We see from figure 1 that this configuration is stable only 
up to very small fluctuations. The
true minima is at $<\chi^9>=0$.  One the other hand turning on a 
non-Abelian vev $<\chi^9> = \sigma_3$  is a true minima. It correponds
to spliting up one ${\tilde D}7$-brane into a D6-brane/anti-D6-brane
pair and the other into an anti-D6-brane/D6-brane pair, i.e. there are
equal numbers of D6-branes and anti-D6-branes at each end point.

\subsection{Quantum Field Theory}

Lastly let us consider the one-loop effective potential for the scalars
by treating \ref{BI7} as a quantum field theory. If $m_i=0$ then the 
Bose-Fermi degeneracy ensures that the effective potential vanishes at 
one-loop. However since the interactions are not supersymmetric we
do not expect a cancellation at higher loops. For $m_i \ne 0$ 
it is not hard to calculate the effective potential for a
scalar vev $<\phi_I> = v_I\sigma_3$ \cite{LS}. 
The result has the classic Coleman-Weinberg
form  given in figure 3. There we included several plots which correspond
to adding various mass renormalisations since, in the absence of
supersymmetry, there appears to be no principle to exclude them. 
\EPSFIGURE{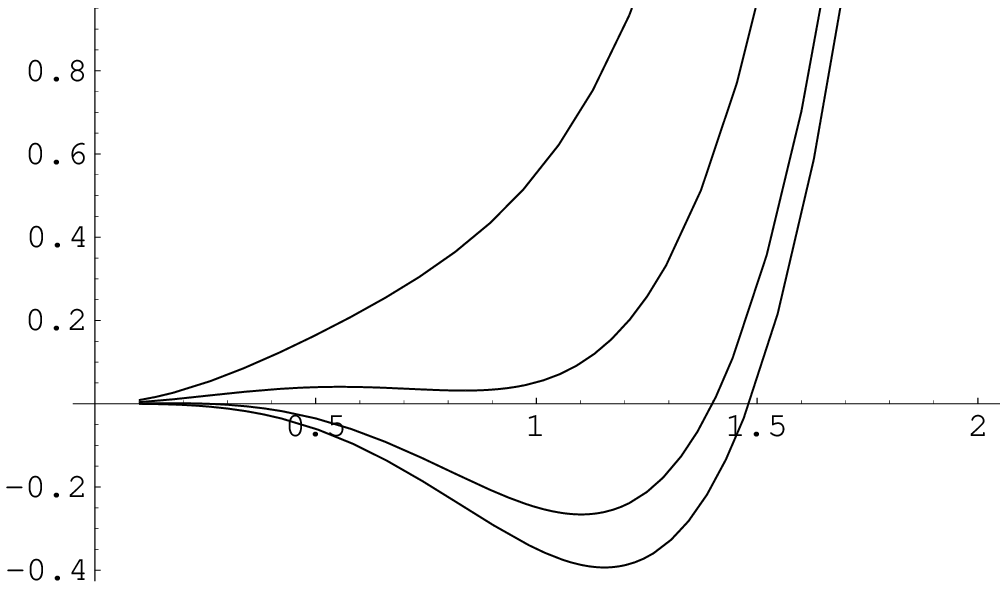,width=3cm}{The Coleman-Weinberg form for
  $V_{eff}$}

Since the $v_I$ parameterise the relative separation of the branes,
$V_{eff}(v_I)$ is the potential between two branes as a function of
their distance. So we see that for small separations there is a
repulsive force between the branes (at least for small or zero values
of the renormalised mass), but
it has a stable minimum and is attractive for large separations.

We should, however, question the validity of the
field theory approximation. Generally speaking the one-loop
approximation in quantum field theory is only valid for large $v_I$
but in our case this is where the masses of the low energy fields
become comparable to the string scale $\alpha'^{-\frac{1}{2}}$
\cite{LS}  and so the low energy approximation is no longer valid. Therefore
the region where the force is attractive should not be trusted. Indeed
we know from the full string theory that the one loop force is repulsive
everywhere \cite{GS}. Although
the small $v_I$ region is strictly speaking not reliable in
quantum field theory one can use renormalisation group methods to
gain a reliable form. Since we do find a repulsive force there, 
in agreement with \cite{GS}, we expect that the basic form for
the effective potential is unaltered by renormalisation group 
techniques.

Now let us return to the case of $m_i=0$ where there are non-Abelian
flat directions and we consider the two cases: $<\chi^9> = v_0 I$
and $<\chi^9>=v_3\sigma_3$. In either case the one loop effective
potential $V_{eff}(v_0)$ or $V_{eff}(v_3)$ again takes on the classic
Coleman-Weinberg form of figure 3 \cite{LS}. However again we can not
trust the large $v_0, v_3$ region. We see from figures 1
and 2 that string theory predicts a non-zero renormalised mass.

In summary it would appear that the effective potential for 
${\tilde D}p$-branes provides an good, and hopefully helpful,
description at the classical level. However 
the one-loop corrections that appear in the quantum field theory,
are only valid for small scalar 
vev's and depend on a priori undetermined mass counter terms which 
are, in turn, fixed in string theory. 

\section*{Acknowledgements}

The work of N.D.L. has been  supported by the EU grant
ERBFMRX-CT96-0012,  the PPARC grant PA/G/S/1998/00613 
and also a PPARC Advanced Fellowship.

\end{document}